\documentclass{elsart}

\usepackage{epsfig}

\usepackage{amssymb}

\begin{document}

\begin{frontmatter}



\title{The (2+1)-dimensional gravastars}

\author[label1]{Farook Rahaman}\footnote{Corresponding author.}\ead{rahaman@iucaa.ernet.in},
\author[label2]{Saibal Ray}\ead{saibal@iucaa.ernet.in},
\author[label3]{A. A. Usmani}\ead{anisul@iucaa.ernet.in},
\author[label4]{Safiqul Islam}\ead{sofiqul001@yahoo.co.in}

\address[label1]{Department of Mathematics, Jadavpur University,
Kolkata 700 032, West Bengal, India.}
\address[label2]{Department of Physics, Government College of Engineering and Ceramic
Technology, Kolkata 700 010, West Bengal, India}
\address[label3]{Department of Physics, Aligarh Muslim University,
Aligarh 202 002, Uttar Pradesh, India}
\address[label4]{Department of
Mathematics, Jadavpur University, Kolkata 700 032, West Bengal,
India.}

\begin{abstract}
We propose a new model of a {\it gravastar} in $(2+1)$ anti-de
Sitter space-time. This new three dimensional configuration has
three different regions with different equations of state: [I]
Interior: $0 \leq r < r_1$, $ \rho = -p$ ; [II] Shell: $r_1 \leq r
< r_2$, $ \rho = p$; [III] Exterior: $r_2 < r $, $ \rho = p =0$.
The outer region of this gravastar corresponds to the exterior
$(2+1)$ anti-de Sitter space-time, popularly known as the BTZ
space-time. Like BTZ model, $\Lambda$ is taken to be negative,
which at the junction turns out to be positive as required by
stability of gravastar and mathematical consistency. After
investigating the Interior space-time, Shell and Exterior
space-time we have highlighted different physical features in
terms of Length and Energy, Entropy, and Junction conditions of
the spherical distribution. It is shown that the present model of
charge-free gravastar in connection to the exterior $(2+1)$
anti-de Sitter space-time or the BTZ space-time is non-singular.
\end{abstract}


\end{frontmatter}

\section{Introduction}
Very recently, we proposed a charged (3+1)-dimensional gravastar
admitting conformal motion \cite{Usmani2011} in the frame work of
Mazur and Mottola model of a chargeless gravastar
~\cite{Mazur2001,Mazur2004}. The model implies that the space of a
gravastar has three different regions defined by different
equations of state (EOS) as: (i) interior:$\: 0<r\leq r_1$,
$p=-\rho$, (ii) shell:$\: 0<r_1<r<r_2$, $p=+\rho$ and (iii)
exterior:$\: r_2<r$, $p=\rho=0$. This provides an alternative to
static black holes, however, energy density is found to diverge in
the interior region of the gravastar, which scales like an inverse
second power of its radius. This behaviour makes the model
singular at $r=0$.

In order to find a non-singular alternative to black holes, it is
worth exploring non-singular gravastars in the (2+1)-dimensions
which offer clarity to study concepts of gravity. Thus, we study
the work of Mazur and Mottola with chargeless gravastar in
(3+1)-dimensions \cite{Mazur2001,Mazur2004} under the
(2+1)-dimensional space-time. We show that the (2+1)-dimensional
neutral gravastars do exist without any curvature singularity at
the origin, which may be considered as an alternative to BTZ black
holes as presented by Ba\~{n}ados, Teitelboim and Zanelli (BTZ)
\cite{BTZ1992}.

It is known that in (2+1) dimensional space-time, Newtonian theory
can not be obtained as a limit of Einstein's theory. In other
words, general relativity in (2+1) dimensions has no Newtonian
limit and no propagating degrees of freedom \cite{Barrow1986}.
Also, gravastar structure can not be obtained from Newtonian
gravity. But it is argued that Einstein's general theory of
relativity admits gravastar structure that has three different
regions such as interior, shell and exterior defined by different
equations of state (EOS) and this provides an alternative to
static black holes. The presence of matter on the thin shell is
required to achieve the crucial stability of such systems under
expansion by exerting an inward force to balance the repulsion
from within. For this reason, (2+1) dimensional gravastar
structure have considered here.

Collapsing matter may appear into a final state of
(2+1)-dimensional black hole with no curvature singularity at the
origin. An explicit vacuum solution of (2+1)-dimensional gravity
with nonzero negative cosmological constant obtained by
Ba\~{n}ados, Teitelboim and Zanelli (BTZ) \cite{BTZ1992} did
demonstrate the existence of such physical systems with an event
horizon and thermodynamic properties similar to (3+1) dimensional
black holes. These are known as BTZ black holes, which are
asymptotically anti-de Sitter. A great number of good reviews on
the subject are available in the literature
~\cite{Carlip1995,Banados1999,Carlip2005}.

For a (2+1)-dimensional gravastar, intermediate region is a
2-dimensional junction instead of a closed spherical shell. Like
BTZ model, $\Lambda$ is taken to be negative, which at the
junction shell turns out to be positive as required for gravastar
and mathematical consistency. It is like requirement of negative
energy at the junction between two separated gravitational sources
to cancel out excess positive energy. The scheme of the
investigation is as follows: In Section II, III and IV
respectively we discuss the Interior space-time, Exterior
space-time and Shell whereas sections V, VI and VII respectively
are devoted to the Length and Energy, Entropy, and Junction
conditions of the spherical distribution.

\section{Interior space-time }
We take the line element for the interior space-time of a static
spherically symmetric distribution of matter in $(2+1)$ dimensions
in the form
\begin{equation}
ds^2 = -e^{2\gamma(r)} dt^2 + e^{2\mu(r)} dr^2 + r^2d\theta^2.
\label{eq1}
\end{equation}

We assume that the matter distribution at the interior of the star
is perfect fluid type, given by
\begin{equation}
T_{ij} = (\rho + p ) u_i u_j + p  g_{ij}, \label{eq2}
\end{equation}

where, $\rho$ represents the energy density, $p$ is  the isotropic
pressure, and  $u^{i}$ is the $3$-velocity of the fluid. The
Einstein's field equations with a cosmological constant ($\Lambda
< 0$), for the space-time described by the metric (\ref{eq1})
together with the energy-momentum tensor given in Eq.~(\ref{eq2}),
yield (rendering $G = c = 1$)
\begin{eqnarray}
\frac{\mu' e^{-2\mu}}{r} &=& 2\pi \rho +\Lambda, \label{eq3} \\
\frac{\gamma' e^{-2\mu}}{r} &=& 2\pi p  -\Lambda, \label{eq4}  \\
e^{-2\mu}\left(\gamma'^2+\gamma''-\gamma'\mu'\right) &=& 2\pi p
-\Lambda, \label{eq5}
\end{eqnarray}
where a `$\prime$' denotes differentiation with respect to the
radial parameter $r$. Combining Eqs.~(\ref{eq3})-(\ref{eq5}), we
get
\begin{equation}
\left(\rho + p\right)\gamma' + p' =0, \label{eq6}
\end{equation}
which is the conservation equation in $(2+1)$ dimensions.

In the interior region I, by the use of the assumption $p=-\rho$
iteratively, the equation (6) yields
\begin{equation}
\rho = constant =\rho_c,~~~(say).
\end{equation}

Let us call this constant as,~$\rho_c = H_0^2/2\pi$ where $H_0$ is
the Hubble parameter with its present day constant value. In other
words, it may be expressed as follows
\begin{equation}
p= -\rho_c.
\end{equation}

We would like to mention that the equation of state $p=-\rho$
(known in the literature as a false vacuum, degenerate vacuum, or
$\rho$-vacuum \cite{Davies1984,Blome1984,Hogan1984,Kaiser1984})
represents a repulsive pressure. In the context of an accelerating
Universe, it is argued by Usmani et al. \cite{Usmani2011} that the
equation of state of the kind $p=-\rho$ may be related to the {\it
$\Lambda$-dark energy}, an agent responsible for the second phase
of the inflation of Hot Big Bang theory
\cite{Riess1998,Perlmutter1999,Ray2007b,Usmani2008,Frieman2008}.

By using the  Eq. (8), one can get the solutions for $\gamma$ and
$\mu$ from the field equations as
\begin{equation}
e^{2\gamma} = e^{-2\mu} = A + (- \Lambda -2\pi \rho_c )r^2,
\end{equation}
where A is an integration constant. Hence, the active
gravitational mass $M(r)$ can be expressed at once in the
following form
\begin{equation}
M(r) = \int_0^R 2 \pi r \rho dr = \pi \rho_c R^2.
\end{equation}
From the above  Eq. (9) we observe that the space-time metric thus
obtained is free from any central singularity.

\section{Exterior space-time}
For the exterior region ($p = \rho = 0$), the BTZ space-time is
\begin{equation}
ds ^2 = - \left(- M_0 - \Lambda r^2\right) dt^2 + \left(- M_0 -
\Lambda r^2\right)^{-1} dr^2 + r^2 d\theta^2.\label{eq22}
\end{equation}
The parameter $M_o$ is the conserved mass associated with
asymptotic invariance under time displacements. This mass is given
by a flux integral through a large circle at space-like infinity.

\section{Shell}
The equation of state in the shell II is $p =\rho$, which
represents a stiff fluid. We note that this type of equation of
state which refers to a Zel'dovich Universe have been selected by
several authors for various situations in cosmology
\cite{Zeldovich1972,Carr1975,Madsen1992} as well as astrophysics
\cite{Braje2002,Linares2004,Wesson1978}.

Within the shell the  Eqs. (3) - (5) can be solved implicitly.
However, it is possible to obtain analytic solution in the thin
shell limit, $ 0 < e^{-2\mu} \equiv h << 1$. We first try to find
implicit solution and later we shall find limiting solution.

\subsection{General case }
By equating  Eqs. (4) and (5), one gets
\[  \gamma'^2+\gamma''-\gamma'\mu' = \frac{\gamma'}{r}\]
which eventually yields
\begin{equation}
 e^\mu = \frac{e^\gamma \gamma'}{r}.
\end{equation}
After a little bit manipulation and calculation, we then get
\begin{equation}
r\gamma'' + 2\Lambda e^{2\gamma} \gamma'^3 + \gamma' =0.
\end{equation}
By the use of any computer programme one can get a solution which,
after exploiting the stiff fluid equation of state $p = \rho$ and
the Eq. (6), is as follows:
\begin{equation}
p = p_0 e^{-2\gamma},
\end{equation}
$p_0$ being an integration constant. Thus, if $\gamma$ is known,
then all parameters can be found in terms of $\gamma$.

\subsection{Thin shell limit}
Let us define $e^{-2\mu} \equiv h(r)$. Then the field  Eqs. (3)
-(6), with $p=\rho$, may be recast in the forms
\begin{equation}
2\Lambda = - \frac{h'}{2r} - \frac{1}{r} \gamma' h,
\end{equation}
\begin{equation}
h =  \frac{r^2} {[(e^\gamma)']^2}.
\end{equation}

Now, it is possible to obtain an analytical solution in the thin
shell limit, $ 0 < e^{-2\mu} \equiv h << 1$. In this limit, we can
set $h$ to zero on the right hand side of (15) leading to order
and integrate immediately to yield
\begin{equation}
h = B - 2\Lambda r^2,
\end{equation}
where $B$ is an integration constant.

The other unknown functions are
\begin{equation}
e^{2\gamma} =  \left[ C -\frac{1}{2\Lambda} \sqrt{B  -2\Lambda
r^2} \right]^2.
\end{equation}

Also, using (14), one can get
\begin{equation}
p = \rho =  p_0\left[ C -\frac{1}{2\Lambda} \sqrt{B -2\Lambda r^2}
\right]^{-2 },
\end{equation}
$C$ being an integration constant.

For consistency of this thin shell limit solutions, we have the
constant pressure term as
\begin{equation}
p_0= \frac{\Lambda_0 C^2}{2\pi},
\end{equation}
where $\Lambda_0$ is the Einstein's erstwhile cosmological
constant.

\section{Proper length and Energy}
We assume the interfaces at $r=R$ and $r=r_2$ describing the phase
boundary of region I and region III respectively are very close.
That is, $r_2 = R + \epsilon$ with $0< \epsilon < < 1$. The proper
thickness between two interfaces i.e. of the shell is obtained as:
\begin{eqnarray}
\ell &=& \int _{R}^{r_2} \sqrt{e^{2\mu} } dr =
\frac{1}{\sqrt{2\Lambda_0}}\left[\sin^{-1}\left(\frac{r\sqrt{2\Lambda_0}}{\sqrt{B}}\right)
            \right]_{R}^{R+\epsilon}.\label{Eq21}
\end{eqnarray}

If we expand $F(R+\epsilon)$ binomially about $R$ and take first
order of $\epsilon$, then, $F(R+\epsilon) \simeq F(R) + \epsilon
F'(R)$ and our $\ell$ would be $ \ell =
\epsilon/\sqrt{B-2\Lambda_0 R^2}$. Obviously, here the length must
be a real and positive quantity and the constraint should be $B
> 2\Lambda_0 R^2$. Actually, the thickness between two interfaces becomes
infinite for the constant value of $B$ as $B = 2\Lambda_0 R^2$ and
hence this identity is not applicable.

We now calculate the energy $\widetilde{E}$ within the shell only
and we get
\begin{equation}
\widetilde{E}  = 2\pi \int _{R}^{R+\epsilon}   \rho r dr =
\nonumber \\  \left[ \frac{8 \pi {\Lambda_0}^2 C}{p_0
(\sqrt{B-2\Lambda_0 r^2}-2 \Lambda_0 C)} -\frac{4 \pi
\Lambda_0}{p_0} \ln ( 2 \Lambda_0 C - \sqrt{B-2\Lambda_0 r^2} )
\right]_{R}^{R+\epsilon}. \label{Eq22}
\end{equation}

In the similar way, as we have done above, by expanding
$F(R+\epsilon)$ binomially about $R$ and taking first order of
$\epsilon$, we get $\widetilde{E} = \epsilon R/[C-
\frac{1}{2\Lambda_0} \sqrt{B-2\Lambda_0 R^2}]^2$ with the
condition to be imposed $B > 2\Lambda_0 R^2$.

\section{Entropy}
Following Mazur and Mottola \cite{Mazur2001,Mazur2004}, we now
calculate the entropy by letting $r_1=R$ and $r_2=R+ \epsilon$:
\begin{equation}
 S = 2\pi\int _{R}^{R+\epsilon} s(r) r \sqrt{e^{2\mu}}dr.
\end{equation}
Here $s(r)$ is the entropy density for the local temperature T(r),
which may be written as
\begin{equation}
s(r) =  \frac{\alpha^2k_B^2T(r)}{4\pi\hbar^2 } =
\alpha\left(\frac{k_B}{\hbar}\right)\sqrt{\frac{p}{2 \pi  }},
\end{equation}
  where $\alpha^2$ is a dimensionless
constant.

Thus the entropy of the fluid within the shell, via the Eq. (24),
becomes
\begin{eqnarray}
S =  2\pi\int _{R}^{R+\epsilon}
\alpha\left(\frac{k_B}{\hbar}\right) \frac{1}{\sqrt{2 \pi p_0}}
\frac{r dr}{\left[ \sqrt{B-2\Lambda r^2}\left(C-\frac{1}{2\Lambda
} \sqrt{B-2\Lambda r^2}\right)\right]}   \nonumber  \\ = \frac{2
\pi \alpha k_B}{ \hbar \sqrt{2 \pi p_0}  } \left[\ln |2 \Lambda C
-  \sqrt{B-2\Lambda r^2} |
  \right]_{R}^{R+\epsilon}.
\end{eqnarray}

\section{Junction Condition}
The discontinuity in the extrinsic curvature determine the surface
stress energy  and surface tension of the junction surface at
$r=R$. Here the junction surface is a one dimensional ring of
matter. Let, $\eta$ denotes the Riemann normal coordinate at the
junction. We assume $\eta$  be positive in the manifold in region
III described by exterior BTZ spacetime and $\eta$ be negative in
the manifold in region I described by our interior  space-time and
$x^\mu = ( \tau,\phi, \eta )$.

The second fundamental forms associated with the two sides of the
shell \cite{Israel1966,Perry1992,Rahaman2006,
Usmani2010,Rahaman2010b,Rahaman2011} are given by
\begin{equation}K^{i\pm}_j =  \frac{1}{2} g^{ik}
\frac{\partial g_{kj}}{\partial \eta}   \mid_{\eta =\pm 0} =
\frac{1}{2}   \frac{\partial r}{\partial \eta} \left|_{r=R}
 ~ g^{ik} \frac{\partial g_{kj}}{\partial r}\right|_{r=R}.
\label{eq36}
\end{equation}
So, the discontinuity in the second fundamental forms is given as
\begin{equation}\kappa _{ij} =   K^+_{ij}-K^-_{ij}.
\label{eq37}
 \end{equation}
In (2+1) dimensional spacetime, the field equations are derived
\cite{Perry1992}:
\begin{eqnarray} \sigma &=&  -\frac{1}{8\pi}  \kappa _\phi^\phi,\\
\label{eq38} v &=&  -\frac{1}{8\pi}  \kappa _\tau^\tau,
\label{eq39}
 \end{eqnarray}
where $\sigma$ and $v$ are line energy density and line tension.
Employing relevant information into Eqs. (28) $\&$ (29) and
setting $r=R$, we obtain
\begin{equation}\sigma =  -\frac{1}{8\pi R}  \left[ \sqrt{ -\Lambda R^2 - M_0}
+  \sqrt{A+ (- \Lambda -2\pi \rho_c)R^2}\right], \label{eq40}
\end{equation}
\begin{equation}v =  -\frac{1}{8\pi}  \left[ \frac{-\Lambda R}
{\sqrt{ - \Lambda R^2 - M_0}} +\frac{(- \Lambda -2\pi \rho_c)R }
{\sqrt{A+ (- \Lambda -2\pi \rho_c)R^2}}\right]. \label{eq41}
\end{equation}

One can note that the line tension is negative which implies that
there is a line pressure as opposed to a line tension. As expected
in (3+1) dimensional case the energy density is negative in the
junction shell. In our configuration, the thin shell i.e. region
II contains ultra-relativistic fluid obeying $p=\rho$ as well as
discontinuity of second fundamental form provides some extra
surface stress energy and surface tension of the junction
interface. These two non-interacting components of the stress
energy tensors are characterizing features of our non-vacuum
region II.\\

\section{Conclusion}
In the neutral gravastar in connection to the exterior $(2+1)$
anti-de Sitter space-time (asymptotically the BTZ space-time) we
have presented a stable and non-singular model. The three
dimensional configuration of this model has three different
regions with different equations of state: an Interior with
geometric and physical structure $0 \leq r < r_1$, $ \rho = -p$ ;
a Shell with $r_1 \leq r < r_2$, $ \rho = p$, and an Exterior with
$r_2 < r $, $ \rho = p =0$.

However, to obtain a realistic picture, following BTZ model
$\Lambda$ is taken to be negative, which at the junction turns out
to be positive as required for the structure of gravastar. As
mentioned in the Introductory part that Ba\~{n}ados, Teitelboim
and Zanelli \cite{BTZ1992} have obtained a unique black hole
solution in the form of point mass with negative cosmological
constant that has a horizon and the radius of curvature $ =
\sqrt(-\Lambda)$ provides the necessary length scale. However, for
$\Lambda \geq 0$ a cosmological type horizon exist with a naked
singularity. The concept of gravastar is to search configuration
which is alternative to black hole. In this paper, we propose the
gravastar configuration which is alternative to BTZ black hole. We
have investigated all these regions and highlighted different
physical features in terms of Length and Energy, Entropy, and
Junction conditions of the spherical distribution.

In this regard we would like to mention that though both the
present work and the one by Usmani et al. \cite{Usmani2011} are
based on the idea of the Mazur-Mottola model but they differ in
two aspects: here conformal motion as well as $3+1$ space-time
have not been adopted. Also, here instead of charge we have
considered a neutral spherical system. Under the $2+1$ dimensional
geometry and non-charged physical structure we have obtained a
non-singular solution for gravastar. However, this demands that
one should investigate a $2+1$ dimensional solution of charged
gravastar.

\section*{Acknowledgments}
\noindent FR, SR and AAU wish to thank the authorities of the
Inter-University Centre for Astronomy and Astrophysics, Pune,
India for providing the Visiting Associateship under which a part
of this work was carried out. Thanks are also due to referee for
the valuable comments which enable us to improve the article
substantially.


\begin{thebibliography}{99}

\bibitem{Usmani2011} A. A. Usmani, F. Rahaman, S. Ray, K. K. Nandi, P. K. F.
Kuhfittig, Sk. A. Rakib, Z. Hasan, Phys. Lett. B 701 (2011) 388.

\bibitem{Mazur2001} P. Mazur, E. Mottola, arXiv:gr-qc/0109035,
Report number: LA-UR-01-5067 (2001).

\bibitem{Mazur2004} P. Mazur, E. Mottola, Proc. Natl. Acad. Sci. USA 101 (2004) 9545.

\bibitem{BTZ1992} M. Ba$\tilde{n}$ados, C. Teitelboim, J. Zanelli, Phys.
Rev. Lett. 69 (1992) 1849.

\bibitem{Barrow1986} J. D. Barrow, A. B. Burd, D. Lancaster, Class. Quantum Gravit. 3 (1986)
551.

\bibitem{Carlip1995} S. Carlip, Class. Quantum Gravit. 12 (1995) 2853.

\bibitem{Banados1999} M. Ba$\tilde{n}$ados, arXiv:hep-th/9901148, DOI: 10.1063/1.59661
(1999).

\bibitem{Carlip2005} S. Carlip, Class. Quantum Gravit. 22 (2005) R85.

\bibitem{Davies1984} C. W. Davies, Phys. Rev. D 30 (1984) 737.

\bibitem{Blome1984} J. J. Blome, W Priester, Naturwissenshaften 71 (1984) 528.

\bibitem{Hogan1984} C. Hogan, Nature 310 (1984) 365.

\bibitem{Kaiser1984} N. Kaiser and A. Stebbins, Nature 310 (1984) 391.

\bibitem{Riess1998} A. G. Riess et al., Astron. J. 116 (1998) 1009.

\bibitem{Perlmutter1999} S. Perlmutter et al., Astrophys. J. 517 (1999) 565.

\bibitem{Ray2007b} S. Ray, U. Mukhopadhyay, X.-H. Meng, Gravit. Cosmol. 13 (2007) 142.

\bibitem{Usmani2008} A. A. Usmani, P. P. Ghosh, U. Mukhopadhyay, P. C.
Ray, S. Ray, Mon. Not. R. Astron. Soc. 386 (2008) L92.

\bibitem{Frieman2008} J. Frieman, M. Turner, D.
Huterer, Ann. Rev. Astron. Astrophys. 46 (2008) 385.

\bibitem{Zeldovich1972} Y. B. Zel'dovich, Mon. Not. R. Astron. Soc. 160 (1972) 1P.

\bibitem{Carr1975} B. J. Carr, Astrophys. J. 201 (1975) 1.

\bibitem{Madsen1992} M. S. Madsen, J. P. Mimoso, J. A. Butcher, G. F. R. Ellis, Phys. Rev. D 46 (1992) 1399.

\bibitem{Braje2002} T. M. Braje, R. W. Romani, Astrophys. J. 580 (2002) 1043.

\bibitem{Linares2004} L. P. Linares, M. Malheiro, S. Ray, Int. J. Mod. Phys. D 13 (2004) 1355.

\bibitem{Wesson1978} P. S. Wesson, J. Math. Phys. 19 (1978) 2283.

\bibitem{Israel1966} W. Israel, Nuo. Cim. B 44 (1966) 1;
erratum - ibid. 48B (1967) 463.

\bibitem{Usmani2010} A. A. Usmani, Z. Hasan, F. Rahaman, Sk. A.
Rakib, S. Ray, P. K. F. Kuhfittig, Gen. Relativ. Gravit. 42 (2010)
2901.

\bibitem{Rahaman2010b} F. Rahaman, K. A. Rahman, Sk. A Rakib, P. K. F.
Kuhfittig, Int. J. Theor. Phys. 49 (2010) 2364.

\bibitem{Rahaman2006} F. Rahaman et al., Gen. Relativ. Gravit. 38
(2006) 1687.

\bibitem{Rahaman2011} F. Rahaman, P. K. F.
 Kuhfittig, M. Kalam, A. A. Usmani, S. Ray, Class. Quantum Gravit. 28 (2011) 155021.

\bibitem{Perry1992} G. P. Perry, R. B. Mann, Gen. Relativ. Gravit. 24 (1992) 305.


\end{thebibliography}
\end{document}